\documentclass{aastex}
\usepackage{emulateapj5}
\usepackage{apjfonts}
\usepackage{epsf}
\usepackage{timesfonts}

\journalinfo{The Astrophysical Journal}

\def\afe{{\cal A}_{\rm Fe}}

\def\ka{{K$\alpha$}~}

\def\rg{R_{\rm G}}

\def\taues{\tau_{\rm es}}

\def\asta{\stackrel{\circ}{\rm A}}

\begin{document}

\title{The Additional Line Component within Iron K$\alpha$ Profile in MCG-6-30-15:
Evidence for Blob Ejection?}

\author{Jian-Min Wang\altaffilmark{1}, 
        Jin-Lu Qu\altaffilmark{1} 
	\& Sui-Jian Xue\altaffilmark{2}}

\altaffiltext{1}{Laboratory for High Energy Astrophysics, Institute of High Energy
Physics, The Chinese Academy of Sciences, Beijing 100039, P.R.  China}
\altaffiltext{2}{National Observatories of China, The Chinese Academy of Sciences, 
Beijing 100020, P.R. China}

\slugcomment{Accepted for Publication in The Astrophysical Journal}
\shorttitle{Evidence for blob ejection in MCG-6-30-15}
\shortauthors{Wang et al.}

\begin{abstract}
The {\em EPIC} data of MCG-6-30-15 observed by {\em XMM-Newton} were analyzed 
for the complexities of the iron K$\alpha$ line. Here we report that the 
additional line component (ALC) at 6.9 keV undoubtedly appears 
within the broad iron K$\alpha$ line profile at high state 
whereas it disappears at low state. These state-dependent behaviors 
exclude several possible origins and suggest an origin 
of the ALC in matter being ejected from the vicinity of the black 
hole. At the low state, the newborn blob ejected from the accretion 
disk is so Thomson thick that hard X-ray is blocked not to ionize 
the old ones, leading to the disappearance of the ALC. When the 
blob becomes Thomson thin due to expansion,  the hard X-ray will 
penetrate it and ionize the old ones, emitting the ALC at the high 
state. The blob ejection is a  key to switch on/off the appearance 
of the ALC.
\keywords{galaxies: active - galaxies: Seyfert: individual: MCG-6-30-15}
\end{abstract}

\section{INTRODUCTION}
MCG-6-30-15 (redshift $z=0.00775$, Fisher et al. 1995) is an ideal laboratory 
in the sky to test how to accrete matter onto a supermassive black hole.
The profile of the iron K$\alpha$ line as a key probe to the general 
relativistic effects have provided an essential clue to understand the details 
of the gas dynamics and radiation mechanism around the black hole. It has been
recognized (see a recent review of Reynolds \& Nowak 2003): 1) the strong general 
relativistic redshifts (including Doppler effects) (Tanaka et al. 1995; Fabian 
et al. 1995; Fabian et al. 2002; Turner et al. 2002; Fabian et al. 2003; Vaughan 
\& Fabian 2004); 2) energy extraction from the spinning black hole (Blandford 
\& Znajek 1977; Wilms et al. 2001; Fabian et al. 2002).

The complexities within the iron K$\alpha$ line profile have been shown
by the appearance of several narrow components. The 6.4 keV
narrow component quite often appears in Seyfert 1 galaxies
(Yaqoob et al. 2000). Turner et al. (2002) show evidence for
the narrow components in NGC 3516, indicating the gravity-controlled
redshifts. Evidence for an absorption edge at 8.7 keV
has been found in Mrk 766, suggesting the presence of ejecta
moving at a velocity of 15000 km/s (Pounds et al. 2003). 
The long time {\em XMM-Newton} observations ($\sim 325$ ks) unambiguously 
presented an additional line component (ALC) at 6.9 keV (Fabian et al.
2002; Fabian \& Vaughan 2003). It is very interesting that the ALC 
did not appear in the observation of 108th revolution (Wilms et al. 2001).
However, the origin of the ALC remains open.  How does this component
appear? Is it related to the blob ejection from the vicinity of
supermassive black hole?

The above questions are the main goals of the present paper.
We show that the ALC is state-dependent on 2.5-10keV X-ray
flux in MCG-6-30-15. This strongly implies it may link with blob 
ejection from the accretion disk.

\section{SPECTRAL ANALYSIS AND RESULTS}
Three sets (revolution 301, 302, 303) of MOS and pn data were taken from
the {\em XMM-Newton Science Archive}. We re-analyzed the data for the 
properties of the ALC with the standard {\tt SAS} pipeline and {\tt XSPEC}. 
Following Fabian et al. (2002), the spectral responses used in the present
paper were {\tt m1\_medv9q19t5r5\_all\_15.rsp} for
MOS-1 (and similarly for MOS-2) and {\tt epn\_sw20\_sdY9\_medium.rmf} for the pn
in the present paper. 
As argued by Fabian et al.(2002) and Fabian \& Vaughan (2003), ``there remain 
problems with the model for charge transfer inefficiency for pn small-window 
mode data'' and  ``it is not clear whether the MOS or the pn 
data give a more accurate description of the shape of the broad iron line". 
Both MOS and pn data were therefore analyzed in the present paper.
It is found that the final results are in agreement with each other 
quantitatively. We give a detail description of MOS data reduction.

The low and high states for pn data are distinguished by a critical count rate 
${\cal R}_{\rm pn}=16$ cts/s in Fabian et al. (2002). This corresponds to
a MOS count rate ${\cal R}_{\rm MOS}=6$ cts/s as shown by the MOS-1/2 light 
curves in Figure 1. The pn light curves have been given in Fabian et al.(2002). 
During the observations of {\em XMM-Newton}'s orbit 108, the 2-10keV fluxes 
were just at the low state (Wilms et al. 2001). The spectra at
the high and low states were produced, respectively. We noted that 
MCG-6-30-15 was in a high state for a large fraction of the period in 
orbit 301, 302, 303, but the length ($\sim 130$ks) at the low state is long 
enough for our goals. The final results of the present work are not very 
sensitive to the critical count rates ${\cal R}_{\rm MOS}$ and 
${\cal R}_{\rm pn}$.

The spectra were fitted with {\tt XSPEC} by the following models: 6.4keV 
narrow component, Compton reflection model, {\tt Laor} profile (Laor 1991) 
and a Gaussian profile of the ALC. In Laor's model, the emissivity of iron 
K$\alpha$ line is assumed $\varepsilon_{\rm K\alpha} \propto R^{-k}$, where 
{\em R} is the radius of the disk and {\em k} is the index. The hydrogen 
column density is fixed at $N_{\rm H}=4.06\times 10^{20}$cm$^{-2}$ (Dickey 
\& Lockman 1990). The inclination angle of the disk ($\Theta$) was fixed at 
28.4$^{\circ}$, and the iron abundance is 3 solar as in Fabian et al. 
(2002) at both the low and high states. At the low state, the best fit 
without the 6.9 keV component is given in Table 1. The errors are quoted 
at the 90\% confidence level. Including this component
gave a slightly poorer fit ($\Delta \chi^2=+0.1$), excluding the presence 
of the ALC at the low state.  The upper 
limits for the equivalent width and luminosity are also given in Table 1. 
At the high state, $\Delta \chi^2=-15.9$ was obtained with a significance
larger than 99\% after we added 6.9 keV ALC. The ALC width 

\begin{figure*}[t]
\centerline{\includegraphics[angle=-90,width=16cm]{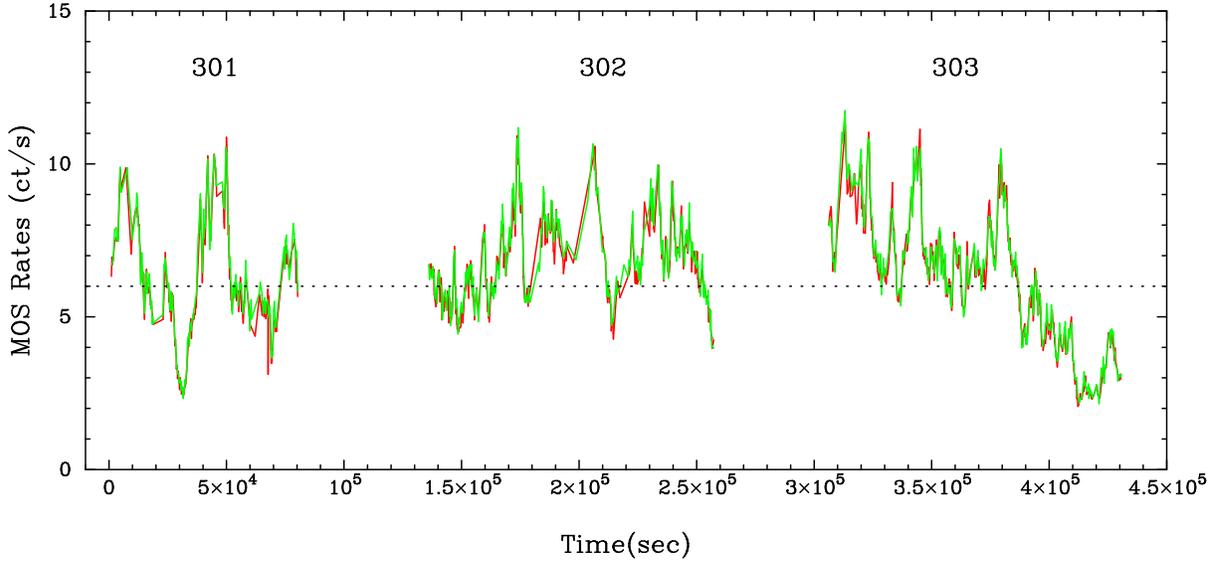}}
\figcaption{The light curves of MCG-6-30-15 by 325ks
{\em XMM-Newton} observations. The red curve is MOS-1 data and
green MOS-2 data.}
\label{fig1}
\end{figure*}

\figurenum{2}
\centerline{\includegraphics[angle=-90,width=8cm]{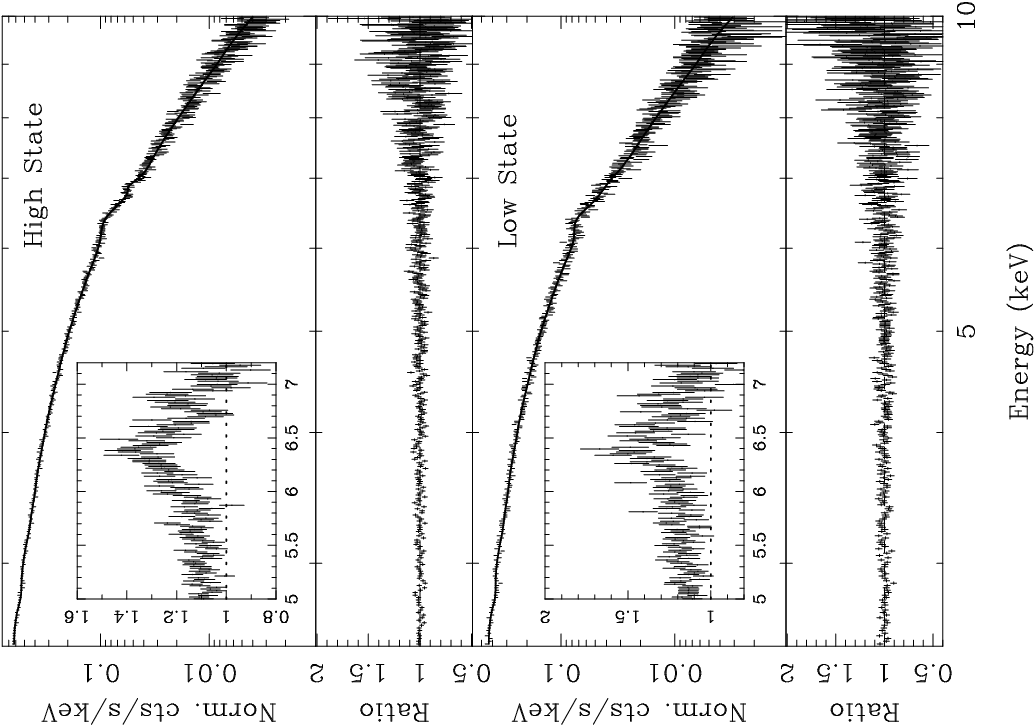}}
\figcaption{The combined MOS spectrum ({\tt addspec} is used to add the MOS-1/2 spectra) 
at the low (lower panel) and the high states (upper panel). The insert panels show a 
close-up of the ratio of the data to the model (the Galactic absorption plus a single 
power law in 2.5-3.0 and 7.5-10 keV) around the iron line region. It is clearly showed 
that the ALC appears at the high state whereas it disappears at the low state.}
\label{fig2}
\vglue 0.45cm

\noindent 
is consistent with zero, but has an upper limit $\sim 70$eV from pn data. We have done 
the ${\cal F}$-test to 
find the significance of the ALC changes from the low state to the high state. 
We found that the statistic value ${\cal F} =5.69$ and the probability $p=7.82 \times 10^{-4}$ 

\figurenum{3}
\centerline{\includegraphics[angle=-90,width=8cm]{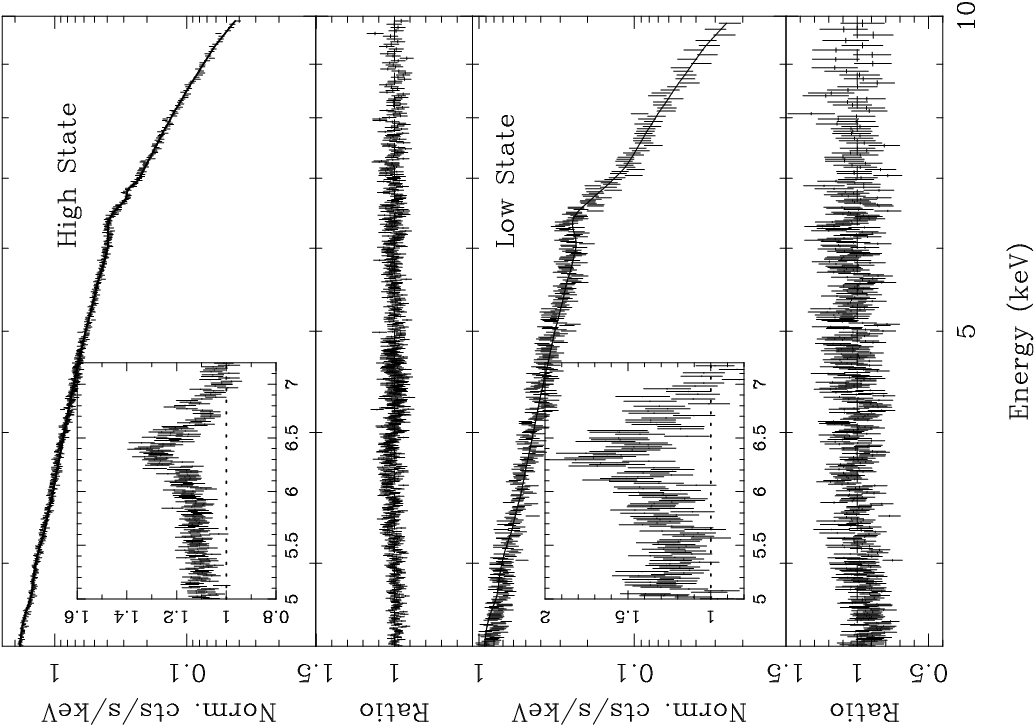}}
\figcaption{The same as in Fig. 2, but for the combined pn spectrum.}
\label{fig2}
\vglue 0.2cm

\newdimen\digitwidth
\setbox1=\hbox{0}
\digitwidth=\wd1
\catcode`"=\active
\def"{\kern\digitwidth}

\def\mc{\multicolumn}
\begin{table*}[t]
\begin{center}
\footnotesize
{\sc Table 1 The Results of fits to the combined MOS-1/2 and pn data}
\vskip 0.3cm
\begin{tabular}{rclrrrcrrrrr} \hline \hline
          & &               &\mc{3}{c}{MOS-1/2}& &\mc{3}{c}{pn data}\\ \cline{4-6} \cline{8-10}
States    & &               & High~~~               & & Low~~~~~              & & High ~~~            & & Low ~~~~\\ \hline
ALC       & &$E_{\rm A}$(keV)&$6.86_{-0.29}^{+0.03}$& & 6.9 (fixed)           & &$6.81_{-0.01}^{+0.05}$ & &6.9 (fixed)\\
          & &$EW$(eV)       &$19.5\pm 7.5$          & & $<12.0^*$             & &$20.0\pm 4.4$        & &$<$11.0    \\
          & &$\sigma$(eV)   &$3.0_{-3}^{+344}$      & & ......                & &$3.0_{-3.0}^{+66.6}$ & &...        \\
          & &$L_{\rm ALC}^1$&$0.85\pm 0.33$         & & $<0.41^*$             & &$0.83\pm 0.18$    & &$<$0.35    \\
          & &               &                       & &                       & &                     & &           \\
Broad \ka & &$E$(keV)       &$6.28_{-0.08}^{+0.09}$ & &$6.98_{-0.34}^{+0.16}$ & &$6.32_{-0.09}^{+0.03}$& &$6.60_{-0.12}^{+0.01}$\\
          & &$EW$(eV)       &$239_{-30}^{+81}$      & &453$_{-67}^{+60}$      & &$256\pm 19 $         & &$499_{-80}^{+40}$ \\
          & &$k$            &$3.48\pm 0.16$         & &$4.30_{-0.53}^{+0.23}$ & &$3.15{\pm 0.08}$     & &$3.73\pm 0.02$  \\
          & &$r_{\rm in}$   &$1.99_{-0.76}^{+0.31}$ & &$1.78_{-0.54}^{+0.19}$ & &$2.85_{-0.19}^{+0.14}$& &$1.40_{-0.17}^{+0.08}$ \\
          & &               &                       & &                       & &                     & &   \\
Narrow \ka& &$EW$(eV)       &$89\pm 10$             & &$80\pm 18$             & &$43.7\pm 5.3$           & &$106_{-19}^{+7}$ \\
          & &$\sigma$(eV)   &$176_{-25}^{+31}$      & &$137_{-49}^{+84}$& &$133\pm 25$          & &$170.0_{-50}^{+80}$ \\
          & &               &                       & &                       & &                     & &                      \\
Continuum & &$\Gamma$       &$1.83\pm 0.01$         & &$1.77_{-0.02}^{+0.03}$ & &$2.03\pm 0.01$       & &$1.89_{-0.05}^{+0.04}$ \\
          & &Fluxes$^2$     &3.17                   & &2.32                   & &3.28                 & & 2.42         \\
          & &               &                       & &                       & &                     & &             \\
$\chi^2/dof$& &             &454.8/488              & &  436.1/489            & &479.0/571            & & 249.6/319\\  \hline
\end{tabular}
\parbox{4.3in}
{\noindent\footnotesize
Note: $^*$ upper limit\\
\indent ~~~~$^1$ The ALC luminosity in unit of $10^{40}$ ergs/s ($q_0=0.5$ and
$H_0=75$~km~s$^{-1}$~Mpc$^{-1}$ are used)\\
\indent ~~~~$^2$ The flux is in 2.5-10 keV in the unit of $10^{-11}$erg~cm$^{-2}$~s$^{-1}$.}
\end{center}
\end{table*}
\vglue 0.2cm

\noindent
(the significance is $1 - p$). Though the upper limit of the 
ALC EW at the low state is close to the lower limit of EW at the high state, 
the difference in luminosity of the ALC between the high and
low state shows the variability of the ALC.
The best fittings are shown by Figure 2 for MOS-1/2 data. It is quite sure that the ALC 
appears at the high state and disappears at the low state. 

The same procedures have been done for the pn data. Figure 3 shows the pn data and the best fitting. 
The results are given in Table 1. We found the EW of the ALC at the high state is obviously stronger 
than that at the low state. The narrow 6.4 keV component shows a significant change in pn data,
especially at the high state, it becomes weaker. The central energy of the broad K$\alpha$ line in pn
data at the low state is lower than that in MOS data, but we should note that the error bar of MOS
data is larger than that of pn data.
We found that the results of pn data are quite similar to that of MOS. 

From Table 1, the broad K$\alpha$ line is enhanced at the low state whereas it 
is weakened at the high state. This property is generally consistent with a long
{\em ASCA} observation (Iwasawa et al. 1996). The ALC has an opposite 
response of the iron K$\alpha$ line to the continuum. 
An extensive analysis of iron K$\alpha$ line has been made in Fabian \& Vaughan (2003). They
found two-component model can explain the variability of iron K$\alpha$ line: a constant (or 
less variable) component and a highly variable power-law component. They suggest that   
the less variable component drives the variability of the K$\alpha$ line. It remains open 
whether there is a relation between the highly variable power-law component and the ALC. 
We will investigate this issue in future.

It has been noted that the central energy of the broad K$\alpha$ line shifts 
to $6.98_{-0.34}^{+0.16}$ keV in MOS data, which perhaps overlaps with the ALC.
The ALC could be thus hidden by the broad K$\alpha$ line at the low state.
We relaxed the ALC energy to explore whether it shifts from the high to 
low states, but we did not get any significantly improved results.
The {\tt XSPEC} fitting only presents the upper limits for its equivalent
and zero width as seen in Table 1. It is very difficult to
separate this component, but the fact should be robust that the
ALC intrinsically weakens (significantly) or disappears at the
low state. Though the current MOS data does not allow us to separate
the ALC from the broad iron K$\alpha$ line at the 
low state, the pn data clearly shows the broad K$\alpha$ line does not overlap the
ALC. The properties of the ALC set strong constraints 
on its origins.

\section{ORIGINS OF THE ALC}
In this section we explore possible originations of the ALC based on
the state-dependent properties. Fabian et al. (2002) ruled out the origin in 
strong iron-L absorption by comparing with soft X-ray spectrum (Lee et al. 2001). 
Here we exclude others and suggest that the ALC originate from the 
matter being ejected from the vicinity of the black hole.


First, recombination lines from a cooling optically thin
plasma blobs ejected from the nucleus (Wang et al. 2000a, Turner et al. 2002, 
Fabian et al. 2002) unlikely appears as the additional line component.  
The K recombination line luminosity is $L_{\rm K}=(4\pi/3)R_p^3n_e^2\varepsilon_{\rm K}$ 
from a hot plasma with radius $R_p$ and number density $n_e$, where $\varepsilon_{\rm K}$ 
is the emissivity of the K line. The Thomson depth is $\taues=n_eR_p\sigma_{\rm Th}$, 
we have $L_{\rm K}=(4\pi/3)\varepsilon_{\rm K}R_p\taues^2/\sigma_{\rm Th}^2$,
where $\sigma_{\rm Th}$ is the Thomson
cross section. For iron hydrogen-like atomic ions, the highest
emissivity $\varepsilon_{\rm K}=1.4\times 10^{-24}\afe$ of Fe XXVI
$\lambda 1.78 \asta$, Fe XXV $\lambda 1.84 \asta$, Fe XXV $\lambda 1.86 \asta$ etc. 
(Raymond \& Smith 1977), the dimension of the hot plasma 
\begin{equation}
R_p \ge 7.5\times 10^{16}L_{\rm K,40} \tau^{-2}_{\rm es,-1}\afe^{-1}~~    (\rm cm),  
\end {equation}
and number density 
\begin{equation}
n_e \le 2.0\times10^6 L^{-1}_{\rm K,40} \tau^3_{\rm es, -1}\afe    (\rm cm^{-3}),  
\end {equation}
are required, where $\taues= 0.1\tau_{\rm es,-1}$, $L_{\rm K,40}=L_{\rm K}/10^{40}$erg/s
and $\afe$ is iron abundance in unit of sun. Thomson depths should be
$\taues\le0.1$, otherwise the recombination line will be greatly broadened by 
Comptonization inside the hot plasma (Pozdniakov et al. 1979). The cooling 
timescale of hot plasma is of
\begin{equation}
t_{\rm ff}\approx 30.0L_{\rm K, 40}T_8^{1/2}\tau^{-3}_{\rm es, -1}\afe^{-1}~~ (\rm yr). 
\end {equation}
Such a long timescale implies that the recombination line should be constant within at 
least the observational interval ($\sim 300$ ks). The ALC should be independent to 
the states of 2-10kev fluxes. This directly conflicts with the observed properties of
the ALC. Moreover,
if it were the recombination line due to the cooling plasma, as a constant line emission,
it should be enhanced at the low state and weakened at the high state
since a constant line flux becomes stronger in an underlying weak continuum.
The disappearance of this component at the low state 
does not favor this origination. 
We note that absorption will be harder to detect at the low state, but more importantly, 
the warm absorber could well be less ionized in the low state, leaving no 
absorption. In order to rule out variability of the absorber, the RGS data should 
be investigated in future.

Second, it is implausible that the ALC originate from a
region of disk photoionized by a flare likely due to a magnetic
field reconnection as suggested for NGC 3516 (Turner
et al. 2002). A multiwavelength study of MCG-6-30-15
showed FWHM(H$\beta)=2400\pm 200$~km/s, luminosity at 5100 $\asta$,
$L(5100 \asta) = 2.53\times 10^{42}$erg/s and the bolometric luminosity
$L_{\rm bol}\approx 4.0 \times 10^{43}$erg/s (Reynolds et al. 1997).
The size of the broad line region of H$\beta$ is $R_{\rm BLR}=2.51$ lt-d from
the empirical reverberation mapping relation 
$R_{\rm BLR}=32.9\left[L(5100 \asta)/10^{44}{\rm erg s^{-1}}\right]^{0.7}$lt-d 
(Kaspi et al. 2000). The mass of the black hole in MCG-6-30-15 is then estimated 
$M_{\rm BH} = (2.1\pm 0.3)\times 10^6 M_\sun$ from the viral formulation
$M_{\rm BH}=1.5 \times 10^5(R_{\rm BLR}/{\rm lt-d})
\left[{\rm FWHM(H\beta)}/10^3{\rm km~s^{-1}}\right]^2M_{\odot}$. 
The analysis of {\em XMM-Netwon} light curves showed $M_{\rm BH}\sim 10^6M_\sun$ for 
the accordance 
with the break frequency $f_{\rm br}\approx 10^{-4}$Hz in its power 
spectrum (Vaughan et al. 2003), which is in good agreement with that 
from empirical reverberation relation. We thus have the accretion rate 
$\dot{m}=\dot{M}/\dot{M}_{\rm Edd}\approx 0.14$, where 
$\dot{M}_{\rm Edd}=1.3\times 10^{24}m_6\eta_{-1}^{-1}$(g/s), 
$m_6=M_{\rm BH}/10^6M_{\sun}$, and  the 
accretion efficiency $\eta_{-1}=\eta/0.1$. For the simplicity, we 
still use the standard 
accretion model to estimate the parameters of the iron K line region 
(Novikov \& Thorne 1973).

A necessary condition of the ionization parameter $\xi\ge 500$
is required for a photoionized plasma on disk by flares to emit
6.9 keV, where $\xi=4\pi F_X/n_H$, $F_X$ is the ionizing X-ray flux
and $n_H$ the hydrogen density (Matt, Fabian \& Ross 1993).
We assume a conversion efficiency ($\eta_X$) of gravitational energy 
into X-ray, and have $F_X=\eta_X F$, where 
$F=3M_{\rm BH}\dot{M}f(R)/(8\pi R^3)$ is the flux of the
dissipated gravitational energy and $f(R)$ is the general relativistic
correction (Novikov \& Thorne 1973). This condition gives a
critical radius for a flare 
$r_c\le 16.2\alpha_{-1}^{2/9}\eta_X^{2/9}(\xi/500)^{-2/9}\dot{m}_{-1}^{2/3}$
(in the units of $R_{\rm G}=GM_{\rm BH}/c^2$)
from the standard disk, where the viscosity of the disk $\alpha_{-1}=\alpha/0.1$
and $\dot{m}_{-1}=\dot{m}/0.1$. The parameter
$\eta_X$ is generally thought to be close to the unit in corona model 
(Haardt \& Maraschi 1991). This critical radius is insensitive to the 
parameters $\alpha$ and $\eta_X$. The width of a line emitted at the radius
$r_c$ is given by
$\sigma=\Delta E/E\approx \sin \Theta/(2r_c)^{1/2}$ for a Schwarzschild black hole 
(Gerbal \& Pelat 1981), where $\Theta$ is the viewing angle,
giving a minimum width of 0.58 keV 
for $\Theta=28.4^{\circ}$. This width is much broader than the observed. 
Thus the origin of the ALC in a highly ionized plasma by a large flare 
on the disk should be ruled out.

Additionally, the Doppler shift due to Keplerian rotation around a 
Schwarzschild black hole (or far away from its center of a Kerr black 
hole) might give rise to the appearance of the ALC. The  
Lorentz factor and velocity of Keplerian rotation are given by 
$\gamma=[(r-2)/(r-3)]^{1/2}$ and $\beta=(r-2)^{-1/2}$ in the unit of light 
speed (see eq.5.4.4a,b on page 423 in Novikov \& Thorne 1973). Then the Doppler 
factor reads ${\cal D}=(r-3)^{1/2}\left[(r-2)^{1/2}\mp \sin \Theta\right]^{-1}$
for approaching and receding direction (Novikov \& Thorne
1973), respectively. If the 6.9 keV line is due to Doppler
blue shift from 6.4 keV, it corresponds to a radius 
$r\approx 22.4$ for $\Theta=28.4^\circ$. The timescale of Keplerian 
rotation is $t_{\rm Kep}\approx 1.0\times 10^3m_6(r/10)^{3/2}$ sec.
The length of the present observations
has several hundred orbits of the Keplerian rotation. Symmetrically
there should be an emission line at 5.56 keV due
to Doppler redshift. We added a Gaussian 

\figurenum{4}
\centerline{\includegraphics[angle=-90,width=8cm]{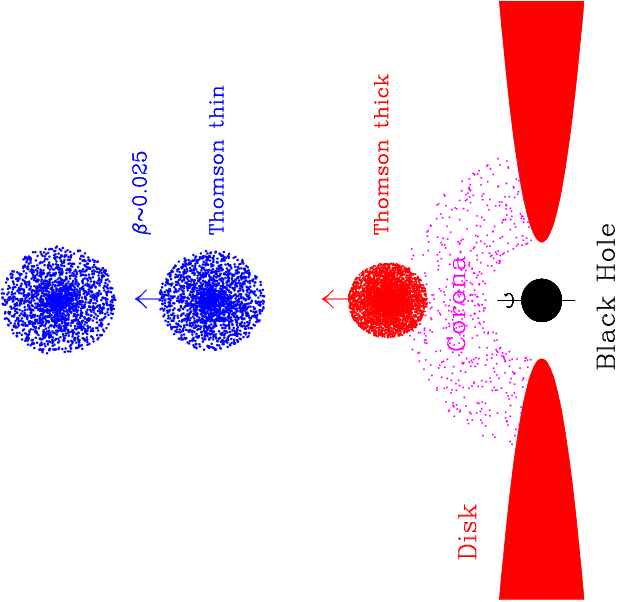}}
\figcaption{ The illustrating model proposed in this paper. The newborn blob
ejected from the vicinity of SMBH is Thomson thick, which prevents the old
blobs from the hard X-ray of the corona. The ALC then disappears at the low
state. The magnetic field pressure inside the blob drives the blob to expand and
become Thomson thin after a period $\sim 10^4$ sec. The hard X-ray can penetrate
the blobs, which are emitting the ALC at the high state. The velocity of the
ejected blobs is $\beta = \upsilon_{\rm b}/c\approx 0.025$. The figure is 
not scaled.}
\label{fig2}
\vglue 0.9cm

\noindent
line at 5.56 keV to
test the plausible existence of this pair line, we found a very
slight improvement $\Delta \chi^2=-0.01$. Such a line does not appear
in the current observational data. The situation in MCG-6-30-15 is thus 
quite different from the case in NGC 3516, which
shows two pairs of narrow emission lines around 6.4 keV due
to Doppler shifts (Turner et al. 2002).

Finally, we have to consider the possibility of origin in the ejected
blobs from the vicinity of SMBH based on the state-dependent behavior 
of the ALC. 

There is some direct evidence for blob ejection to take place
at the low state in 3C 120 (Marscher et al. 2002), and microquasars
(Livio et al. 2003). The ejected clouds are suggested
for the 8.7 keV absorption edge in Mrk 766, but the ejection mechanisms 
are highly uncertain (Pounds et al. 2003). In a new model of jet formation,
the transients of the accretion disk is due to jet production. The disk
is at the low state when the blob ejection takes place there. The transition is 
caused by most likely a large scale magnetic (Livio et al. 2003), which 
can be produced by the disk dynamo (Tout \& Pringle 1996). The angular
momentum transportation, production of the large scale magnetic field
and blob ejection are entangled with each 
other. As a consequence of the angular momentum conservation, the 
ejected blob may whirl around the black hole spin axis. This would result
in more complicate and sophisticated effects. This is beyond the scope of the
current data. We do not discuss this further in the present paper.
The following is devoted to an outline of the model of blob's 
evolution to explain the observation.

We introduce a ratio $q=\dot{M}_{\rm eje}/\dot{M}$, the fraction of
the ejection to the accretion rate of the disk. The value of $q$ is highly 
unknown, but it should be a significant fraction of the accretion
rate estimated from the observations. In the case of MCG-6-30-15, we have
a very crude estimate of $q\sim 1-F_{\rm low}^X/F_{\rm high}^X\sim 0.2$ from 
the Table 1. The ejection lasts a period of
the low state $\Delta t$ (Marscher et al. 2002; Livio et al. 2003).
We assume that the ejected blob has an initial radius 
$R_{b,0}=r_{b,0}R_G$, the density of the blob is $\rho_0=\Delta M_{b}/
\frac{4}{3}\pi R^{3}_{b,0}$, where 
$\Delta M_{b}=\dot{M}_{\rm eje}\Delta t=q\dot{M}\Delta t$. 
The Thomson depths of the newborn blob is given 
by $\tau_{\rm es,0}={\kappa_{\rm es}}{\rho}_{0} R_{b,0}$, namely,
\begin{equation}
\taues=4.69\times 10^2q_{-1}\dot{m}_{-1}\Delta t_4r_b^{-2}m_6^{-1},
\end{equation}
where electron scattering opacity $\kappa_{\rm es}=0.34$, $q_{-1}=q/0.1$, 
and $\Delta t_4=\Delta t/10^4$sec. Such a Thomson
thick blob will block the X-ray to illuminate the blobs,
which have been ejected before this newborn blob. The ALC from the
ejected blobs will be quenched at the low state.

However, the magnetic field pressure inside the newborn
blob is driving it to expand. The expanding velocity can be roughly 
estimated by $\frac{1}{2}\rho_{0}\upsilon_{b}^{2}=B^2/8\pi$, we have
\begin{equation}
\upsilon_{\rm b}=7.5\times 10^7B_4r_0^{1/2}m_6(\taues/500)^{1/2}~~{\rm (cm~s^{-1})}, 
\end{equation}
where the magnetic field $B_4 =B/2.5\times 10^4$G ($\sim$
the equipartition value). The ejection velocity
of the blob can be estimated $\upsilon_{\rm eje}\approx 0.025c$ from the
Doppler blue shift ${\cal D}_{\rm max}=6.86/6.7\approx 1.023$
for the viewing angle $\Theta=28.4^\circ$.
The sub-relativistic velocity is consistent
with the very weak radio emission in MCG-6-30-15
(Ulvestad \& Wilson 1984). The opening angle of the blobs
will be $\theta_b=R_b/H_b\approx \upsilon_b/\upsilon_{\rm eje}\approx 0.1$,
where $H_b$ is the distance
of the ejected blob from the black hole. The ionization parameter
of the ejected blobs outside the corona is given by
\begin{equation}
\xi=1.1\times 10^2\theta_{\rm b,0.1}L_{X,42}h_{b,3}^{-1}m_6^{-1}\tau_{\rm b,es}^{-1}, 
\end{equation}
where $h_{b,3}=H_b/10^3\rg$ and $\theta_{b,0.1}=\theta_b/0.1$. Setting $\xi=500$,
we found a critical height
\begin{equation}
H_b^c/\rg=2.21\times 10^2\tau_{\rm b,es}^{-1}m_6^{-1}L_{X,42}\theta_{b,0.1},
\end{equation}
above which the iron is emitting $6.4\sim 6.7$keV K$\alpha$ line and below which iron 
ions are fully ionized. We obtained another critical height of the blob,
\begin{equation}
H_b^{\taues}/\rg=2.17\times 10^2\theta_{b,0.1}^{-1}(q_{-1}\dot{m}_{-1})^{1/2}\Delta
t_4^{1/2}m_6^{-1/2},
\end{equation}
when the Thomson depth $\tau_{\rm b,es}=1$. It is found that $H_b^c\approx H_b^{\taues}$.
This implies that it is the time for the ejected blob to emit iron K$\alpha$ line when 
the Thomson depth of the blob becomes thin. The newborn blob
then switches on X-ray to ionize the old blobs, leading to a recurrence of the ALC. 
The proposed model is illustrated in Figure 3. The fluorescent K$\alpha$ line luminosity
from one ejected blob is given by (Krolik \& Kallman 1987),
\begin{equation}
L_{\rm K\alpha}\approx 2.34\times 10^{39}\theta_{b,0.1}^2{\cal A}_{\rm Fe,3}\tau_{\rm b,es}
~~(\rm erg~s^{-1}),
\end{equation}
where the optical depth of the K$\alpha$ edge
$\tau_{\rm K\alpha}\approx 2.0{\cal A}_{\rm Fe,3}\tau_{\rm b,es}$, 
${\cal A}_{\rm Fe,3}=\afe/3$, the mean fluorescent yield $Y\approx 0.6$,
the observed X-ray spectrum is used as the illuminating source 
$L_{\varepsilon}=4.1\times 10^{41}(\varepsilon/\varepsilon_{\rm K\alpha})^{-0.83}$
erg~s$^{-1}$~keV$^{-1}$ at the high state, and line energy $\varepsilon_{\rm K\alpha}=6.7$~keV.
The observed ALC luminosity is of $8.5\pm 3.3\times 10^{39}$erg~s$^{-1}$, which needs about 4
blobs. 

The lifetime of the ejected blob is estimated by 
$\tau_{\rm es}=0.01{\tau}_{\rm es,-2}$ due to expansion when the blob contributes less 
to the observed ALC. It yields the final radius 
$R_{b}/R_G=2.17\times 10^{2}(q_{-1}\dot{m}_{-1})^{1/2}
(\Delta t_{4}/{m_{6}})^{1/2}\tau_{\rm es,-2}^{-1/2}$ from equation (4), 
implying a lifetime $t_{\rm blob}=R_b/\upsilon_{b}$, i.e.
\begin{equation}
t_{\rm blob}\approx 5.0\times 10^{5} B_{4}^{-1}r_{b,0}^{1/2}\tau_{\rm es,-2}^{-1/2}~~ (\rm sec).
\end{equation} 
The typical lifetime is long enough to observe the process that the newborn blob 
switches on/off the old blobs to emit the ALC. The proposed model focuses
on the ALC behaviors, but the ejected blobs are undergoing
acceleration by radiation pressure and the ionization parameter
is changing (Chelouche \& Netzer 2001). A detail model of
blobs undergoing a magnetic field-driven expansion is in preparation.

A Gaussian profile of the ALC is assumed in the present paper. Thermal broadening 
is quite small ($\Delta E/E \approx 4.3\times 10^{-4}T_6^{0.5}$, where
$T_6=T/10^6$K, then $\Delta E \approx 3$eV at the ALC energy $E\sim 6.9$keV
for a $10^{6}$ K plasma). The profile of the iron K line 
from the ejected blobs is mainly controlled by several other factors:
1) velocity distribution of the ejected blob which may be influenced
by the interaction between the blob and its surroundings
(for example,the medium of wind from the disk);
2) redshift due to gravity of the black hole in the center; 
3) the changes of ionization degrees of the blobs with distance from the center.
The detail profiles of a line from the jet around a black hole 
has been done by Wang et al.(2000b). However the present
EPIC data poorly gives the information on the detail profile of the ALC.

Interestingly, Pounds et al. (2003) found an absorption at
8.7keV in Mrk 766, which shifts with flares. This implies 
the line-of-sight absorption in Mrk 766. They suggest a
'cloud' (ejected from the disk) model for this absorption and predict 
an emission feature significantly contributed at 6.7keV. 
Most recently, 
Turner et al. (2003) found a narrow component at 5.7 keV in Mrk 766, 
suggesting the emission from the decelerated blobs (Wang et al. 2000a).
The pn data do not show absorption edge feature above 8.0 keV at both
the low and high states in MCG-6-30-15. The case in MCG-6-30-15 may be
different from Mrk 766. These features should be investigated in future.

It should be mentioned that a highly Doppler blueshifted Fe
K emission line appeared in two quasars PKS 2149-306 (Yaqoob et al. 1999) 
and CXOCDFS J033225.3-274219 (Wang et al. 2003). PKS 2149-306 is a
radio-loud quasar and the jet has a velocity of $0.75c$.
The later has a velocity of $0.6-0.7c$. The jets in the two 
objects have much faster velocities than the blob's  in MCG-6-30-15.
If the model for the ALC suggested is correct in the present paper, it would be
of interests in what reason is responsible for the difference of the ejection 
velocities. A future study on the comparison of MCG-6-30-15 with the two quasars
might shed light on this difference. We need further observations with higher 
energy resolution and longer monitoring time to probe the physical processes taking
place around the supermassive black hole.

\section{CONCLUSIONS}
We show in the present work that the appearance of the ALC
depends on the states of the source. At the low state, this component
disappears whereas it appears clearly at the high state.
We argue that this state-dependent behaviors of ALC strongly indicate the
origin in the blobs ejected  from the innermost region of the accretion
disk. A simple model for the state-dependent behaviors of the ALC has been outlined 
in this paper. The newborn blob is initially Thomson thick and switches off the X-ray 
from hot corona to ionize the old blobs, giving rise to the ALC disappears. With
expansion of the newborn blob driven by the magnetic field, it becomes Thomson thin
and switches on X-ray to ionize the blob  and then ALC appears. Future mission,
such as {\em Astro-E II}, may resolve the detail (profile and possible different
components) of the ALC with high energy resolution.
This would help to explore the origin of the ALC and its implications.
Additionally, monitoring MCG-6-30-15 is desired to test variabilities of the ALC and the
relation with the continuum so 
as to understand the process of the blob ejection. This process is an intermediate state 
between relativistic radio jets and 
the large outflow velocities observed in Seyfert galaxies. It is
believed that the X-ray emission lines would trace the processes of the blob ejection. 

\acknowledgements{The authors acknowledge a referee for the careful and helpful comments 
significantly improving the paper. J.M.W. is grateful to T. J. Turner, S. Komossa, R. Staubert 
and K. Werner  for interesting discussion. S.M. Jia and Y. Chen are greatly thanked for their 
help in {\em XMM} data reduction. This research is financed by the Special Funds for Major
State Basic Research Projects and supported by Grant for Distinguished Young Scientist from
NSFC.}

\end{document}